%% file: odyssey20a_submission.tex
\documentclass[a4paper]{article}
\usepackage{Odyssey2020}
\usepackage{epsfig,amssymb,amsmath}
\usepackage{pifont}
\usepackage{multirow}
\usepackage{makecell}
\usepackage{listings}
\usepackage{xcolor}

\usepackage[utf8]{inputenc}

\definecolor{codegreen}{rgb}{0,0.6,0}
\definecolor{codegray}{rgb}{0.5,0.5,0.5}
\definecolor{codepurple}{rgb}{0.58,0,0.82}
\definecolor{backcolour}{rgb}{0.95,0.95,0.92}
 
\lstdefinestyle{mystyle}{
    backgroundcolor=\color{backcolour},   
    commentstyle=\color{codegreen},
    keywordstyle=\color{magenta},
    numberstyle=\tiny\color{codegray},
    stringstyle=\color{codepurple},
    basicstyle=\ttfamily\footnotesize,
    breakatwhitespace=false,         
    breaklines=true,                 
    captionpos=b,                    
    keepspaces=true,                 
    numbers=left,                    
    numbersep=5pt,                  
    showspaces=false,                
    showstringspaces=false,
    showtabs=false,                  
    tabsize=2,
    basicstyle=\fontsize{7}{9}\ttfamily
}
 
\title{Delving into VoxCeleb: environment invariant speaker recognition}
 \name{Joon Son Chung*, Jaesung Huh*\thanks{\hspace{-12pt}* These authors contributed equally to this work.}, Seongkyu Mun}
 \address{Naver Corporation, South Korea\\
 {\small \tt joonson.chung@navercorp.com }}

\begin{document}
%\ninept
%
\maketitle
\begin{abstract}
Research in speaker recognition has recently seen significant progress due to the application of neural network models and the availability of new large-scale datasets. There has been a plethora of work in search for more powerful architectures or loss functions suitable for the task, but these works do not consider what information is learnt by the models, apart from being able to predict the given labels.

In this work, we introduce an environment adversarial training framework in which the network can effectively learn speaker-discriminative and environment-invariant embeddings without explicit domain shift during training. We achieve this by utilising the previously unused `video' information in the VoxCeleb dataset. The environment adversarial training allows the network to generalise better to unseen conditions. The method is evaluated on both speaker identification and verification tasks using the VoxCeleb dataset, on which we demonstrate significant performance improvements over baselines.

\end{abstract}

% -------------------------------------------------------------------------

\section{Introduction}
\label{sec:intro}

Deep learning has been pushing the state-of-the-art in many fields of research over the recent years. These architectures can simultaneously learn feature representation and decision framework from large labelled datasets, removing the need to handcraft features for any given problem. 
Such progress has been facilitated by the availability of large-scale datasets, such as ImageNet~\cite{Deng09} for image classification, Labeled Faces in the Wild~\cite{Huang07} for face recognition and VoxCeleb~\cite{Nagrani17} for speaker recognition.
However, the weakness of data-driven approaches is that it is not possible to define what information is learnt by the models during the training process -- whether it is the useful information or undesirable biases that are present in the dataset.

In speaker recognition, the challenge comes down to the ability to separate the voice characteristics and the environments in which the person's voice is recorded. The VoxCeleb dataset contains recordings from diverse but finite environments for each speaker, making it possible for the model to overfit to the environment as well as the voice characteristics. In order to prevent this, we must look beyond classification accuracy as the only learning objective.

In this paper, we propose a new framework for learning effective speaker embeddings at the same time as removing undesirable sources of variation such as environment information. This work is inspired by domain adaptation approaches~\cite{ganin2016domain,liu2016coupled,tzeng2015simultaneous,tzeng2017adversarial} and an extension of this work to bias removal in models~\cite{Alvi18}.
Also of relevance are recent works~\cite{rohdin2019speaker, bhattacharya2019adapting,bhattacharya2019generative} that have used adversarial training in speaker recognition for adaptation between distinct domains. A more detailed overview of these work will be given in Section~\ref{subsec:related}.

In contrast to the previous work on domain adaptation, our model is trained to be invariant to environments and recording conditions {\em without} explicit domain shift or the use of domain annotation during training. 
The model is trained on the VoxCeleb dataset alone and there is no supervisory requirement beyond what is provided in the dataset.
The network trained using the proposed framework generalises better to both unseen samples of seen speakers for speaker identification and to unseen speakers for speaker verification, as well as to unseen environments.

The paper is organised as follows. In Section~\ref{sec:framework}, we discuss the adversarial learning framework that allows speaker representations to be trained whilst removing environment information without explicit labels. Section~\ref{sec:exp} describes the trunk architectures for the network and the dataset used for training. In Section~\ref{sec:res}, we demonstrate that the speaker recognition networks trained using the proposed framework yields significant improvements over baselines. The experiment on `replayed' VoxCeleb dataset show that the gains are more pronounced in environments not seen during training. We also probe the trained network to find that much of the environment information has indeed been removed from the embedding.

\subsection{Related works}
\label{subsec:related}

Although this paper focuses on adversarial training, there has been a long history of research on methods to train noise and environment robust speaker embeddings, ranging from pre-processing to data augmentation. Each of these will be described in the following paragraphs.

\vspace{5pt}\noindent{\bf Traditional methods.}
% In the early stage of speaker recognition when Gaussian mixture model(GMM) based universal background model (UBM) was widely used, various approaches using model adaptation and factorization have been proposed . 
Traditional literature on speaker recognition~\cite{hansen2015speaker} have used speaker-dependent GMM mean components as speaker representation features, which are obtained by adapting the UBM to the speaker's voice. However, the adaptation, usually maximum a posteriori (MAP), adapts not only to speaker-specific characters of speech, but also to channel and other nuisance factors. To address this issue, joint factor analysis~\cite{kenny2008study} and i-vector based approach~\cite{dehak2010front}, which decomposes speaker and channel components using eigenvoice or total variability matrix, have been used. However, in environment robust research \cite{hasan2013maximum,lei2013noise}, these techniques have not been commonly applied to deep learning based methods since they require direct update of model statistics such as mean and covariance of GMM.

As reported in \cite{mclaren2015advances, snyder2018x}, the use of mid-level features of DNN has generally yielded higher performance compared to the i-vector framework, when large training database can be utilised.

\vspace{5pt}\noindent{\bf Pre-processing.}
% Speech enhancement has been proposed as a way to improve the performance of speaker recognition systems by reducing the noise at the input.
% These approaches show significant Signal-to-Noise Ratio (SNR) improvement, but this does not guarantee improvement of speaker recognition performance directly. 
DNN-based speech enhancement has been a popular field of research in the recent years, using generative adversarial networks (GAN)~\cite{soni2018time} or Wave-U-Net~\cite{giri2019attention}.
A recent work has analysed the effects of speech enhancement to speaker recognition~\cite{novotny2019analysis}, and \cite{shon2019voiceid} has proposed joint training speech enhancement network together with speaker recognition system in order to improve its robustness. Whilst there is a difference between the speaker recognition performance and perceptual signal quality, this work shows the potential of the joint training to improve the overall performance.

% These approaches are meaningful in that they focused on improving speaker recognition performance, not just improve quality of enhanced signal. Moreover, it is interesting to point out the difference between the actual speaker recognition performance and signal quality based on human perception. However, these methods have some limitations that an additional training stage needs to be done and overall size of the network becomes heavier.

% \textcolor{red}{Recently, beamforming and feature extraction using multichannel microphones have been proposed for speaker recognition \cite{taherian2019deep, cai2019multi, movsner2018dereverberation}. As the multi-channel approach showed a promising performance improvement in the speech recognition field, a meaningful improvement  can also be found in speaker recognition cases. However, using multiple microphones is difficult to apply in all cases, due to cost and structure issues.}

\vspace{5pt}\noindent{\bf Data Augmentation.}  
Data augmentation has been used widely in machine learning to improve the performance of systems by artificially increasing the amount and diversity of training data. This technique was first popularised in computer vision~\cite{Ciresan12,Krizhevsky12} by using label preserving transformations to augment image data. The method has been adopted for speaker recognition~\cite{snyder2018x} by adding noise~\cite{snyder2015musan} and reverberations~\cite{ko2017study} to the speech signal. This recipe has been used by many of the recent work in the field~\cite{yamamoto2019speaker, qin2019far,zeinali2019but}.

\vspace{5pt}\noindent{\bf Adversarial training.} 
% Another approach is to utilize multi-task learning which is proven to be effective in various fields such as computer vision~\cite{girshick2015fast, kendall2018multi}. A number of researchers tried to apply this method to speaker recognition task. Jati et al. \cite{jati2019multi} introduced DNN-TVM model along with multi-task learning but the purpose of the training scheme is to obtain both correct classification performance and distinctive speaker embeddings. Ding et al. \cite{ding2018mtgan} applied multi-tasking to enhanced triplet loss with generative adversarial mechanism but they didn't target to train environment-invariant speaker recognition network. 
Deep learning approaches have made breakthroughs in performance for many applications of machine learning, but the trained models often generalise poorly to unseen data. Recent works~\cite{ganin2016domain,tzeng2015simultaneous} on domain adaptation have proposed to address this issue by introducing {\em domain confusion loss} in addition to the standard classification loss. The domain confusion loss tries to match the distribution between source and target domains in order to confuse the classification layers, so that the samples from both domains are indistinguishable for the classifier. 

Similar approaches have been adopted in speaker recognition to adapt speaker embeddings to a new domain. \cite{rohdin2019speaker, bhattacharya2019adapting,bhattacharya2019generative} have used domain adaptation training similar to~\cite{ganin2016domain,tzeng2015simultaneous} between languages~\cite{rohdin2019speaker, bhattacharya2019adapting,bhattacharya2019generative} and between datasets~\cite{bhattacharya2019adapting,wang2018unsupervised}. 

Of closest relevance to our work is \cite{Alvi18} that extends the work on domain adversarial training to {\em bias removal} for face recognition task by introducing auxiliary classifiers and confusion losses for undesirable sources of variations (e.g. age, nationality) that the primary classifier (identity) should be invariant to.  \cite{zhou2019training} have replicated similar strategy for speaker recognition task, where the auxiliary classes are the types of noise that have been added to the speech signal. 

The commonality amongst the previous work, on speaker recognition or otherwise, is that they require explicit domain or class labels in the source of variation that the trained embedding should be invariant to, and this must be a well-defined category. In contrast, our method is capable of learning environment-robust embeddings without explicit domain shift or auxiliary class labels.

% -------------------------------------------------------------------------

\section{Learning framework}
\label{sec:framework}

In this section, we propose a training framework that aims to learn a feature representation that encapsulates useful speaker information, whilst being uninformative for undesirable sources of variations such as environment or recording conditions.

An overview of our framework is given in Figure~\ref{fig:overview}.
The speaker network has a single classification loss, which classifies the audio segment into one of 1,211 speakers. The environment network is first trained to determine whether or not the two audio segments are from the same environment, and also used to remove this information from the speaker embedding.

\subsection{Batch formation}

\begin{figure}[!h]
\centering 
\includegraphics[width=0.7\linewidth]{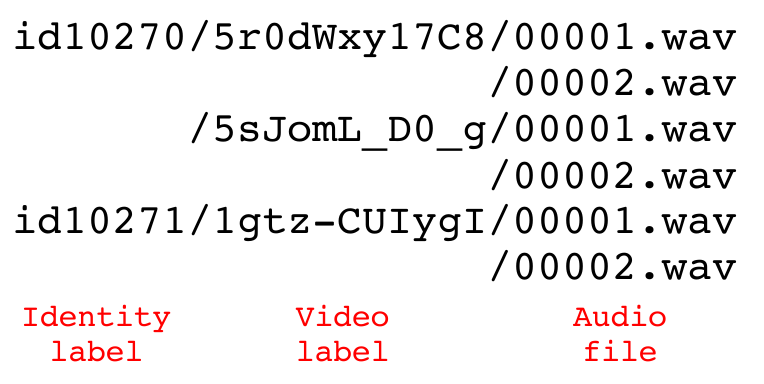}
\caption{Data labels in the VoxCeleb dataset.}
\label{fig:filename} 
\end{figure}

\begin{figure*}[!htb]
\centering 
\includegraphics[width=1.0\linewidth]{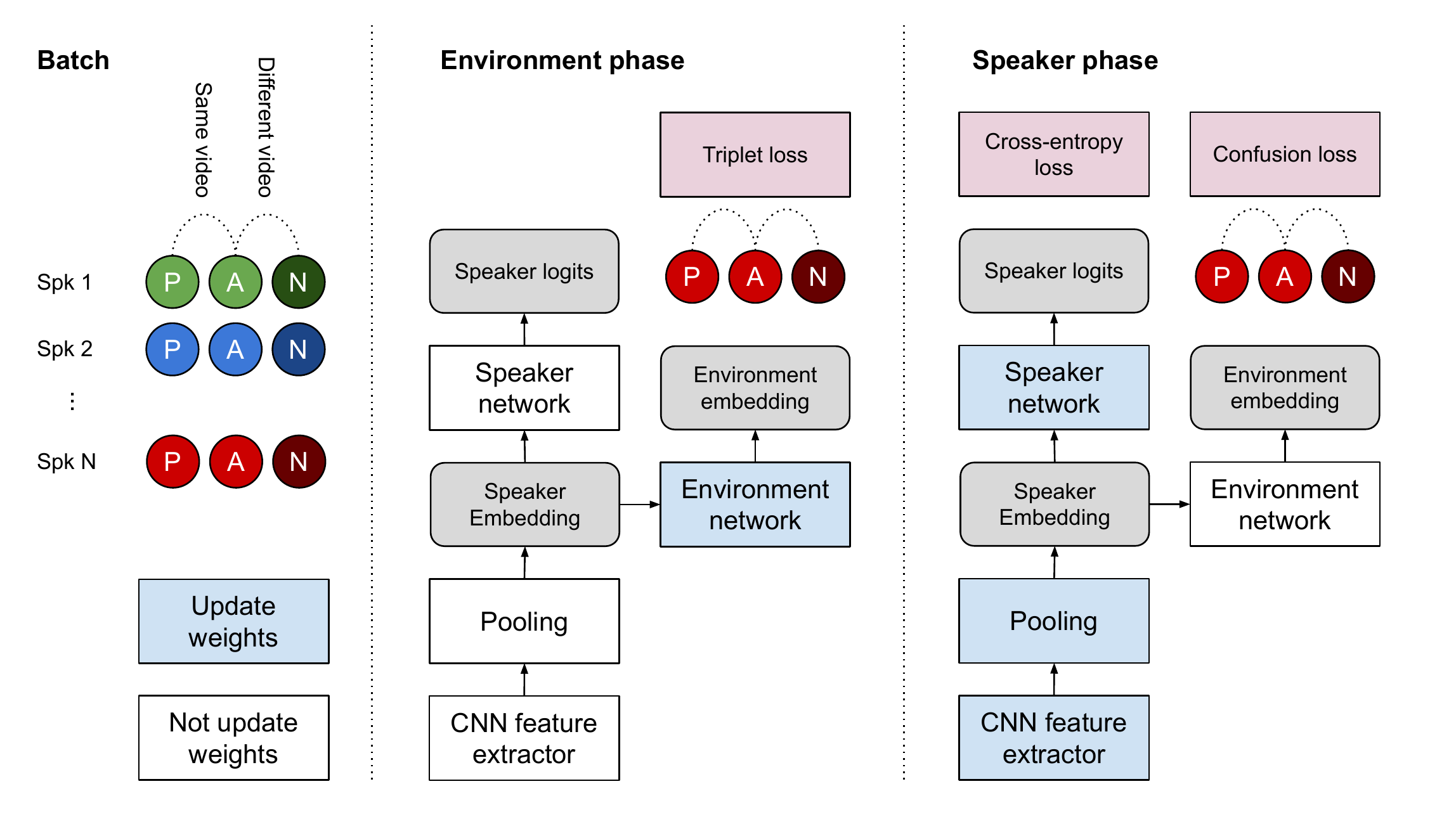}
\caption{Overview of the training strategy. `Confusion loss' minimises the KL divergence between the softmax of the triplet distances and a uniform distribution. {\bf P} (positive) and {\bf A} (anchor) are from the same video, {\bf N} (negative) is from a different video from the anchor. }
\label{fig:overview} 
\vspace{10pt}
\end{figure*}

\begin{lstlisting}[float=*,language=Python,caption={PyTorch-style pseudocode for the training scheme},label={pseudo}]
## Let xa, xp, xn, y be anchor, positive, negative speaker inputs and class labels.
## netcnn is the CNN feature extractor, netspk and netenv are fully connected networks.

optimizer       = optim.SGD([netcnn.parameters(), netspk.parameters()]);
disc_optimizer  = optim.SGD(netenv.parameters());

for xa, xp, xn, y in loader:
    ca, cp, cn      = netcnn.forward(xa), netcnn.forward(xp), netcnn.forward(xn)
    
    # Environment phase
    ea_, ep_, en_   = netenv.forward(ca.detach()), netenv.forward(cp.detach()), netenv.forward(cn.detach())
    disc_loss       = torch.mean(F.relu(torch.pow(l2_dist(ea_, ep_)) - torch.pow(l2_dist(ea_, en_)) + margin))
    disc_loss.backward();
    disc_optimizer.step();
    
    # Speaker phase
    sa, sp, sn      = netspk.forward(ca), netspk.forward(cp), netspk.forward(cn)
    ea, ep, en      = netenv.forward(ca), netenv.forward(cp), netenv.forward(cn)
    triplet_logits  = F.log_softmax(torch.stack((l2_dist(ea, ep),l2_dist(ea, en)),dim=1))
    x               = torch.cat((sa,sp,sn),dim=0)
    loss            = nn.CrossEntropyLoss(x, y.repeat(3)) + alpha * nn.KLDivLoss(triplet_logits, target=uniform)
    loss.backward();
    optimizer.step();
\end{lstlisting}

Each minibatch consists of three 2-second audio segments from $N$ different speakers. Two of the three audio segments from each speaker are from the same video, and the other is from a different video. The two segments can be either from different parts of the same audio clip or another clip from the same YouTube video. The video reference can be inferred from the file path in the VoxCeleb dataset, as shown in Figure~\ref{fig:filename}. Here, the assumption is that the pair of clips from the same video would have been recorded in similar environments, whereas the clips from different videos would have more different channel characteristics.
An example of such input batch is depicted in left hand column of Figure~\ref{fig:overview}.

\subsection{Algorithm}
The algorithm is described by the pseudocode in Listing~\ref{pseudo} and described below. 

The speaker embedding is first extracted by the CNN feature extractor, and pooled over time using one of the two pooling strategies described in Section~\ref{sec:agg}.

\vspace{5pt}\noindent\textbf{Environment phase.}
The environment network is trained to predict whether or not the input audio segments come from the same environment (same video) with a triplet loss, the anchor and a segment from the same video as the anchor forming the positive pair, and the anchor and a segment from a different video forming a negative pair. 
Suppose $x_a, x_p, x_n$ are anchor, positive and negative speaker input representations and $e_a, e_p, e_n$ are the corresponding outputs of the environment network. Then, environment loss $L_e$ is 
\begin{equation}
L_{e} = max(0, || e_a - e_p ||_{2}^2 - || e_a - e_n ||_{2}^2 + \mathit{m}) 
\label{eq:le}
\end{equation}
where $m$ is a margin of triplet loss.  The gradient is back-propagated only to the environment network, so the CNN feature extractor is not optimized during this phase. 

\vspace{5pt}\noindent\textbf{Speaker phase.}
The CNN feature extractor and the speaker recognition network are trained simultaneously using the standard cross-entropy loss. In addition, the confusion loss penalises the network's ability to discriminate between the clips that originate from the same environment (same video) and those from different environments -- this is done by minimising the entropy between the softmax of the triplet distances and a uniform distribution. 
The environment or channel information can be seen as undesirable sources of variations that should be absent from an ideal speaker embedding.

The loss function in this phase is computed as follows. Let $\mathbf{s} = \{s_a, s_p, s_n \}$ be the corresponding outputs of the speaker network and $y$ be the speaker label of $\mathbf{s}$. The loss function $L_{s}$ is
\begin{equation}
    \begin{gathered}
        % d_{x,y} = || x - y ||_{2}^2 \\
        % TL = =\frac{\exp(d_{x_a,p})}{\exp(d_{a,p}) + \exp(d_{a,n})} \\
        p_{dist} = softmax({|| e_a - e_p ||_{2}^2,|| e_a - e_n ||_{2}^2}) \\
        L_{s} = CE(\mathbf{s}, y) + \alpha * KL( p_{dist} || p_{unif}) 
    \end{gathered}
\label{eq:se}
\end{equation} 
where $p_{unif}$ is an uniform distribution. Here, $CE(\mathbf{s},y)$ is the cross entropy loss between speaker logits $\mathbf{s}$ and label $y$ . $KL(p_a||p_b)$ is the KL divergence between distribution $p_a$ and $p_b$.
The extent to which the confusion loss contributes to the overall loss function is controlled by the variable $\alpha$ (e.g. 0, 1, 10, 30).

% -------------------------------------------------------------------------

\section{Models and Experiments}
\label{sec:exp}

This section describes the network architecture, the dataset and details of the experiments.

\input{resnet.tex}

\subsection{CNN feature extractor}
\label{sec:feat}

Experiments are performed on two different architectures and in Table~\ref{table:convnet}.

\vspace{5pt}\noindent\textbf{VGG-M-40.}
The original VGG-M model has been proposed for image classification~\cite{Chatfield14} and adapted for speaker recognition by~\cite{Nagrani17}. Whilst it is not a state-of-the-art network, the network is known for high efficiency and good classification performance. VGG-M-40 is a further modification of the network used by~\cite{Nagrani17}  to take 40-dimensional filterbanks as inputs instead of the 513-dimensional spectrogram, significantly reducing the number of computations. 

\vspace{5pt}\noindent\textbf{Thin ResNet-34.}
Residual networks~\cite{He16} are used widely in image recognition and has recently been applied to speaker recognition~\cite{cai2018exploring,Chung18a,Xie19a}. Thin ResNet-34 is the same as the original ResNet with 34 layers, except with only one-quarter of the channels in each residual block in order to reduce computational cost.

\subsection{Temporal aggregation}
\label{sec:agg}

Since we want the network to be invariant to temporal position but {\em not} frequency, \cite{Nagrani17} has proposed aggregation layers that are fully connected only along the frequency axis. This produces a $1 \times T$ feature map before the pooling layers, described in the following sections.

\vspace{5pt}\noindent\textbf{Temporal average pooling (TAP).} 
The TAP layer simply takes the mean of the features along the time domain. 

\vspace{5pt}\noindent\textbf{Self-attentive pooling (SAP).}
Unlike the TAP layer that equally pools the features over time, \cite{cai2018exploring} introduces a self-attentive pooling layer to pay attention to the frames that are more informative for utterance-level speaker recognition. This is effectively a weighted mean of the features (Equation~\ref{eq:weightedsum}), where the weights $w_t$ are given by Equation~\ref{eq:w1} and \ref{eq:w2} where $W$, $b$ and $\mu$ are learnable matrices or vectors. $x_t$ are utterance-level feature maps along time domain and $e$ is the final speaker embedding.

\begin{equation}
e = \sum_{t=1}^{T} w_t x_t
\label{eq:weightedsum}
\end{equation}
\begin{equation}
h_t = tanh(Wx_t + b)
\label{eq:w1}
\end{equation}
\begin{equation}
w_t = \frac{exp(h_t^T \mu)}{\sum_{t=1}^T exp(h_t^T \mu)}
\label{eq:w2}
\end{equation}

\subsection{Speaker and environment networks}
\label{sec:network}

The speaker network is a linear classifier that consists of a single fully connected layer with the output size equal to the number of speakers (1,211). 

The environment network has two fully connected layers of size 512, each preceded by ReLU activation and batch normalisation.

\subsection{Dataset}
\label{sec:dataset}

We train our models end-to-end on the VoxCeleb1 dataset. It contains more than 150,000 utterances from 1,251 speakers. 
% Development and test sets are fixed, both speaker identification and verification, for the fair comparison. 

For {\bf identification}, we only train on the {\em overlapping part} of the development sets for identification and verification, so that the models trained for identification can be used to evaluate verification. This makes speaker identification a 1,211-way classification task, and the test set consists of unseen utterances of seen speakers during training.

For {\bf verification}, all speech segments from the 1,211 development set speakers are used for training, and the trained model is then evaluated on the 40 unseen test set speakers. The statistics are summarised in Table~\ref{table:dataset}.

\begin{table}[!h]
\centering
\begin{tabular}{ l |r |r }
\textbf{Split} & \textbf{\# Speakers}  & \textbf{\# Utterances}   \\ \hline

Dev (Iden.) &  1,211 & 140,638          \\ 
Test (Iden.) &  1,211 & 7,972          \\ \hline

Dev (Ver.) &  1,211 & 148,610         \\ 
Test (Ver.) &  40 & 4,874          \\ 
\end{tabular}
\vspace{2pt}
\caption{Development and test splits for speaker identification and verification. Note that the identification split is different from that in the original VoxCeleb paper~\cite{Nagrani17}.}
\label{table:dataset}
\end{table}

\input{results.tex}

\subsection{Training details}
\label{sec:training}

\vspace{5pt}\noindent{\bf Input representations.}
During training, we use a fixed length 2 second temporal segment, extracted randomly from each utterance.
Spectrograms are extracted with a hamming window of width 25ms and step 10ms. For ResNet, the 257-dimensional raw spectrograms are used as the input to the network. For VGG networks, 40-dimensional Mel filterbanks are used as the input.
Mean and variance normalisation (MVN) is performed on every frequency bin of the spectrogram and filterbank at utterance-level.
No voice activity detection (VAD) or data augmentation is used in training.

\vspace{5pt}\noindent{\bf Speaker verification.}
The network has been trained for a $n$-way classification task, but the verification task requires a measure of similarity. The final layer in the classification network is replaced with one of output dimension 512, and this layer is re-trained with contrastive loss and hard negative mining. The preceding layers (i.e. the CNN feature extractor) is {\em not} finetuned with the contrastive loss, which is in line with the training procedure of~\cite{Nagrani17,Chung18a}.

\vspace{5pt}\noindent{\bf Implementation details.}
Our implementation is based on the PyTorch framework~\cite{paszke2017automatic} and trained on NVIDIA Tesla P40 accelerators. The network is trained using Stocastic Gradient Descent (SGD) with an initial learning rate of $10^{-3}$, decreasing by a factor of 0.95 every epoch. Batch normalisation~\cite{Ioffe15} is used during training. The training is stopped after 100 epochs or whenever the validation error did not improve for 10 epochs, whichever is sooner.

% -------------------------------------------------------------------------

\section{Results}
\label{sec:res}

In this section, we first compare the performance of our method to baselines, and also probe the network to see if the environment information has been removed from the embedding.

\subsection{Speaker recognition}

The trained network is evaluated on the VoxCeleb1 test set. We sample ten 2-second temporal crops from each test segment, and compute the distances between all possible combinations
($10 \times 10 = 100$) from every pair of segments. The mean of the 100 distances is used as the score. This protocol is in line with that used by~\cite{Chung18a}.

Table~\ref{tab:results} reports results for multiple models used for evaluation. Across both speaker identification and verification tasks, the models trained with the proposed adversarial strategy ($\alpha > 0$) consistently outperform those trained without ($\alpha=0$). 

\subsection{Replay experiment}
\label{sec:replay}

\begin{figure}[!htb]
\centering 
\includegraphics[width=1.0\linewidth]{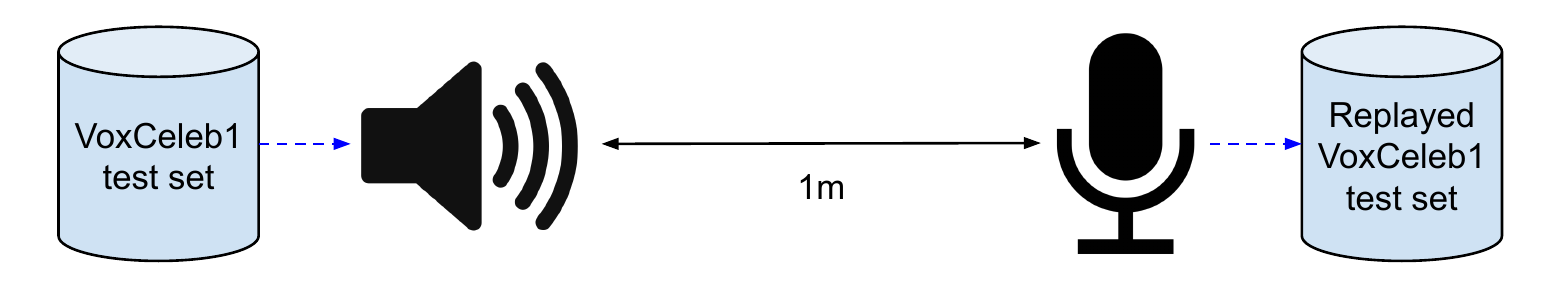}
\caption{Setup for the replay experiment. }
\label{fig:replay} 
\end{figure}

The `replay' experiment measures the performance on the same VoxCeleb test set used in the speaker verification task, but played through a reference speaker and re-recorded using a Jabra SPEAK 510 microphone. This results in a significant change in channel characteristics and deterioration of sound quality. The models are identical to those used in previous experiments, and not fine-tuned on the replayed segments.

The results are shown in the column of Table~\ref{tab:results} named {\bf Replay EER}. The improvement in performance as $\alpha$ increases is more pronounced in this setting (24\% relative improvement in EER for the ResNet model compared to 9\% improvement in the original VoxCeleb dataset), which suggests that the model trained with the proposed adversarial training generalises much better to unseen environments or channels.

\subsection{Analysis on the removal of environment information}
We perform experiments to verify that the adversarial training helps to remove environment information from the embedding. The test list for evaluating the environment recognition consists of 9,486 same speaker pairs, half of which come from the same video and the other half from different videos. We report the results in the right-most column of Table~\ref{tab:results}. 
A lower EER indicates that the network is better at predicting whether or not a pair of audio segments come from the same video.
The results demonstrate that environment recognition performance decreases with the increase of $\alpha$, which shows that the unwanted environment information has indeed been removed from the speaker embedding to an extent.

% -------------------------------------------------------------------------
\section{Conclusion}
\label{sec:con}

In this paper, we have proposed an environment adversarial training framework in which the network can learn speaker-discriminative and environment-invariant embeddings. We have evaluated the proposed method on both speaker identification and verification tasks using the VoxCeleb dataset, on which the performance of the method exceeds that of baselines by a significant margin on an unseen test set. We also probe the trained network to verify that much of the environment information has indeed been removed from the embedding.

\vspace{12pt}\noindent\textbf{Acknowledgements.}
We would like to thank Triantafyllos Afouras, Jisu Choi, Sanghyuk Chun, Hee Soo Heo and Bong-Jin Lee for helpful comments.

% -------------------------------------------------------------------------
\clearpage
\bibliographystyle{IEEEbib}
\bibliography{longstrings,vgg_local,vgg_other,mybib}

\end{document}

%% file: resnet.tex
\begin{table*}[ht]
\centering
\begin{tabular}{ c|c|c }
\textbf{layer name} & \textbf{VGG-M-40}  & \textbf{Thin ResNet-34}   \\ \hline
conv1 & \makecell{$5 \times 7,96$, stride 2 \\ $ 3 \times 3$, max pool, stride $ 1 \times 2$}  & \makecell{$7 \times 7,16$, stride 2 \\ $ 3 \times 3$, max pool, stride 2}  \\   \hline
%\addlinespace[1.5ex]
conv2 &  \makecell{$5 \times 5,96$, stride 2 \\ $ 3 \times 3$, max pool, stride 2}    & $\begin{bmatrix}  3 \times 3, 16 \\ 3 \times 3, 16 \end{bmatrix} \times 3 $, stride 1         \\\hline
%\addlinespace[1.5ex]
conv3 & $3 \times 3,256$, stride 1 & $\begin{bmatrix}  3 \times 3,32 \\ 3 \times 3,32 \end{bmatrix} \times 4 $, stride 2      \\\hline
%\addlinespace[1.5ex]
conv4 & $3 \times 3,256$, stride 1 & $\begin{bmatrix}  3 \times 3,64 \\ 3 \times 3,64 \end{bmatrix} \times 6 $, stride 2        \\\hline
%\addlinespace[1.5ex]
conv5 &  \makecell{$3 \times 3,256$, stride 1 \\ $ 3 \times 3$, max pool, stride 2}    & $\begin{bmatrix}  3 \times 3,128 \\ 3 \times 3,128 \end{bmatrix} \times 3 $, stride 2       \\\hline
%\addlinespace[1.5ex]
fc &  $ 4 \times 1$, 512, stride 1  &  $ 9 \times 1$, 512, stride 1                  \\ 
\end{tabular}
\vspace{2pt}
\caption{Modified VGG-M and ResNet architectures. ReLU and batchnorm layers are not shown.  Each row specifies the number of convolutional filters, their sizes and strides as {\bf size $\times$ size, \# filters, stride}. The output from the fully connected layer is ingested by the pooling layers.}
\label{table:convnet}
\vspace{10pt}
\end{table*}

%% file: results.tex
\begin{table*}[h] 
\begin{center}
\begin{tabular}{ |c | c | c | r | r | r|r|r|r|r|} 
 \hline
 \textbf{Model} &
 \textbf{Pooling} &
 \textbf{{\bf $\alpha$}} &
 \textbf{Iden. T1} &
 \textbf{Iden. T5} &
 \textbf{Ver. EER} &
 \textbf{Replay EER} &
 \textbf{Env. EER} 
 \\ \hline
 VGG-M~\cite{Nagrani17} & TAP & 
           0 & - & - & 7.82\% & - & - \\ \hline
   \multirow{4}{60pt}{~~VGG-M-40} & \multirow{4}{40pt}{~~~~TAP} &
           0 & 67.62\% & 82.90\% & 8.44\% & 15.16\% & {\bf 18.72\%}  \\ 
    &   & 1 & 72.74\% & 87.15\% & 8.15\% & 14.60\% & 20.01\%  \\
    &   & 10 & 75.96\% & 89.48\% & 7.79\% & 14.23\% & 21.93\% \\
    &   & 30 & {\bf 77.70\%} & {\bf 90.29\%} & {\bf 7.61\%}  & {\bf 13.21\%} &  23.87\% \\\hline
   \multirow{4}{60pt}{~~VGG-M-40} & \multirow{4}{40pt}{~~~~SAP} &
           0 & 68.13\% & 83.67\% & 8.02\% & 15.74\%  & {\bf 18.45\%}  \\ 
    &   & 1 & 70.99\% & 85.64\% & 8.31\% & 14.68\% & 18.98\%  \\
    &   & 10 & 76.01\% & 89.53\% & 7.93\% & 14.49\%  & 21.63\% \\
    &   & 30 & {\bf 77.31\%} & {\bf 90.48\%} & {\bf 7.82\%} & {\bf 13.88\%} & 23.55\%  \\\hline
%   \multirow{4}{80pt}{~~~Thin ResNet-34} & \multirow{4}{40pt}{~~~~SAP} & \multirow{4}{40pt}{~~~~~~\xmark} &
%           0 & 67.58\% & 82.30\% & 8.28\% & {\bf 15.39\%} \\ 
%     & &  & 1 & 75.71\% & 88.14\% & 7.73\% & 17.92\% \\
%     & &  & 10 & 80.58\% & 91.98\% & 7.04\% & 20.77\% \\
%     & &  & 30 & {\bf 81.45\%} & {\bf 92.66}\% & {\bf 6.59\%} &  21.70\% \\\hline
   \multirow{4}{80pt}{~~~Thin ResNet-34} & \multirow{4}{40pt}{~~~~SAP} &
           0 & 83.68\% & 92.90\% & 5.71\%  & 13.14\% & {\bf 20.43\%} \\ 
    &  & 1 & 88.34\% & 95.48\% & 5.38\% & 12.41\% & 23.04\%  \\
    &   & 10 & {\bf 89.00\%} & 95.94\% & {\bf 5.26}\% & {\bf 10.56\%}  & 25.74\%  \\
    &   & 30 & {\bf 89.00\%} & {\bf 96.15\%} & 5.37\%  & 10.58\% & 26.38\% \\\hline
\end{tabular}
\end{center}
\caption{
Results on speaker identification and verification tasks. Models for both tasks have been trained {\em only} on VoxCeleb1. 
Note that the test set split for the identification task is different from~\cite{Nagrani17}. 
{\bf T1, T5:} top-1 and top-5 accuracies; {\bf EER:} Equal Error Rate.
}
\label{tab:results}
\vspace{10pt}
\end{table*}